\documentclass{aa}
\usepackage{graphicx}
\usepackage{natbib} 
\bibpunct{(}{)}{;}{a}{}{,}

\begin{document}
 
\title{Testing the Seyfert unification theory: Chandra HETGS observations
of NGC 1068}
 
\author{P. M. Ogle\inst{1} \and T. Brookings\inst{1} \and C. R. Canizares \inst{2} 
        \and J. C. Lee, \inst{2} \and H. L. Marshall \inst{2}}
\institute{UC Santa Barbara, Physics Department, UCSB, Santa Barbara, CA 93106
            \and MIT CSR, Cambridge, MA 02139}

\offprints{P. M. Ogle, \email{pmo@xmmom.physics.ucsb.edu}}
\titlerunning{Testing Seyfert unification in NGC 1068}
\date{Received 15-05-02/ Accepted 04-11-02}
\abstract{  
We present spatially resolved {\it Chandra} HETGS observations of the Seyfert 
2 galaxy \object{NGC 1068}. X-ray imaging and high resolution spectroscopy are
used to test the Seyfert unification theory. Fe K$\alpha$ emission is 
concentrated in the nuclear region, as are neutral and ionized continuum 
reflection. This is consistent with reprocessing of emission from a luminous, 
hidden X-ray source by the obscuring molecular torus and X-ray narrow-line region 
(NLR). We detect extended hard X-ray emission surrounding the X-ray peak in the nuclear 
region, which may come from the outer portion of the torus. Detailed modeling of the
spectrum of the X-ray NLR confirms that it is excited by photoionization and 
photoexcitation from the hidden X-ray source. K-shell emission lines from 
a large range of ionization states of H-like and He-like N, O, Ne, Mg, Al, Si, S, 
and Fe {\sc xvii}-{\sc xxiv} L-shell emission lines are modeled. The emission
measure distribution indicates roughly equal masses at all observed ionization 
levels in the range $\log \xi=1-3$. We separately analyze the spectrum of an off-nuclear 
cloud. We find that it has a lower column density than the nuclear region, and is also 
photoionized. The nuclear X-ray NLR column density, optical depth, outflow velocity, and 
electron temperature are all consistent with values predicted by optical 
spectropolarimetry for the region which provides a scattered view of the hidden
Seyfert 1 nucleus. 

\keywords{galaxies: active -- galaxies: individual: \object{NGC 1068} -- X-rays: galaxies}
}

\maketitle


\section{Introduction}

\object{NGC 1068} (z=0.003793) contains one of the brightest and closest 
Seyfert 2 active galactic nuclei (AGN). It has also served as the exemplar of 
AGN unification for  Seyfert 1 and 2 galaxies \citep{am85}. Broad optical 
lines from a hidden Seyfert 1 nucleus are scattered by electrons in the nuclear
region. It is hypothesized that Seyfert 1 and 2 galaxies are primarily 
distinguished by whether we view the nucleus directly or indirectly by 
scattered and reprocessed light. This model can be tested in many ways using 
high resolution spatial and spectral observations with the {\it Chandra X-ray 
Observatory}.

One of the predictions of the Unified Seyfert Model is the existence of an 
optically thick torus which collimates the ionizing radiation from the 
nucleus and blocks our direct view of the nuclei of Seyfert 2 galaxies. 
Observations of the hard ($E>10$ keV) X-ray continuum of \object{NGC 1068} 
with {\it Beppo-SAX} show that it is dominated by Compton reflection, and 
suggest that the torus is Compton-thick \citep{mgf97}. The attendant narrow, 
large equivalent width Fe K$\alpha$ emission line \citep{kit89,m93,ifm97} most 
likely comes from fluorescence in the Compton-thick torus, with a possible 
contribution from the ionized X-ray narrow-line region (NLR) 
\citep{kk87,m93,mbf96}.  Our high-resolution {\it Chandra} observations allow 
us to constrain the location, size, and kinematics of the Fe K$\alpha$ 
emitting region and investigate the properties of the torus. 

It has been suggested that warm ionized gas ablated from the torus provides
the optical scattering medium \citep{kb86,kk95}. One important implication of 
electron scattering by warm gas in the NLR is that there must be a phase of 
the NLR with temperature $T\sim 10^5$K. This plasma should also reveal itself in the
X-ray band via scattered continuum, recombination emission, and resonance 
scattering.  We make high-resolution measurements of these spectral components
with {\it Chandra} HETGS, and derive strong constraints on the properties of 
the electron scattering region. 

Another important goal is to understand the physical state and dynamics of AGN
outflows. The bi-conical narrow-line region (NLR) seen in many Seyfert galaxies
is one manifestation of these outflows. Optical spectroscopy of the 
\object{NGC 1068} NLR with the {\it Hubble Space Telescope} ({\it HST}) has 
revealed the kinematics of the inner NLR, which appears to be  an outflowing 
wind \citep{ck00}. Interaction of this wind with the radio jet and host galaxy
ISM may also be important \citep{cdg01}.

Recent {\it Chandra} imaging spectroscopy of \object{NGC 1068} 
shows a detailed correspondence between extended X-ray emission and the 
optical narrow-line region \citep{yws01}. Similar extended X-ray nebular
emission is observed in NGC 4151, Mrk 3, and Circinus \citep{o00,skp00,snk01}.
{\it XMM} RGS observations of the soft X-ray spectrum of \object{NGC 1068} show 
the importance of photoexcitation and give the first measurements of the temperature 
and column density of X-ray emitting gas in the NLR \citep{ksb02}. We apply a similar 
analysis to {\it Chandra} HETGS spectra of \object{NGC 1068}, which have 
better resolution and S/N in the medium-hard (0.8-7 keV) band.

The high spatial resolution of {\it Chandra} HETGS enables us to measure 
the spectrum in two locations in the NLR--the central 1.5\arcsec of
the nuclear region, and a cloud $3\arcsec$ NE of the nucleus (subsequently,
 ``NE cloud''). We use these spectra to measure and compare the column 
densities in the two regions. Multi-band X-ray imaging also reveals how the 
X-ray spectrum changes across the NLR. Near-simultaneous observations with 
{\it Chandra} LETGS are presented in a parallel paper by \cite{b02}.

\section{Observations}

We observed \object{NGC 1068} on 2000 December 4-5 (MJD 51882-3) with 
{\it Chandra} HETGS. The total exposure time, after correcting 
for dead time is 47.0 ks. The dispersion axes (Roll$=308.9\degr$, HEG 
PA$=123.7\degr$, MEG PA$=133.6\degr$) are nearly perpendicular to the axis of 
the NLR (PA$\simeq 40\degr$). This maximizes the spectral resolution and 
enables spectroscopy at multiple locations along the extended NLR. The width 
of the nuclear emission region in the dispersion direction for the zeroth 
order image ranges from FWHM$=0.81-0.66\arcsec$ for $E=0.6-3$ keV. This 
corresponds to a smearing of FWHM $=0.015-0.018$ \AA\ over the 6-22 \AA\ range
in addition to the instrumental profile (FWHM$=0.01$, 0.02 \AA\ for HEG, MEG).
  
\begin{figure*}
\centering
\includegraphics[width=17cm]{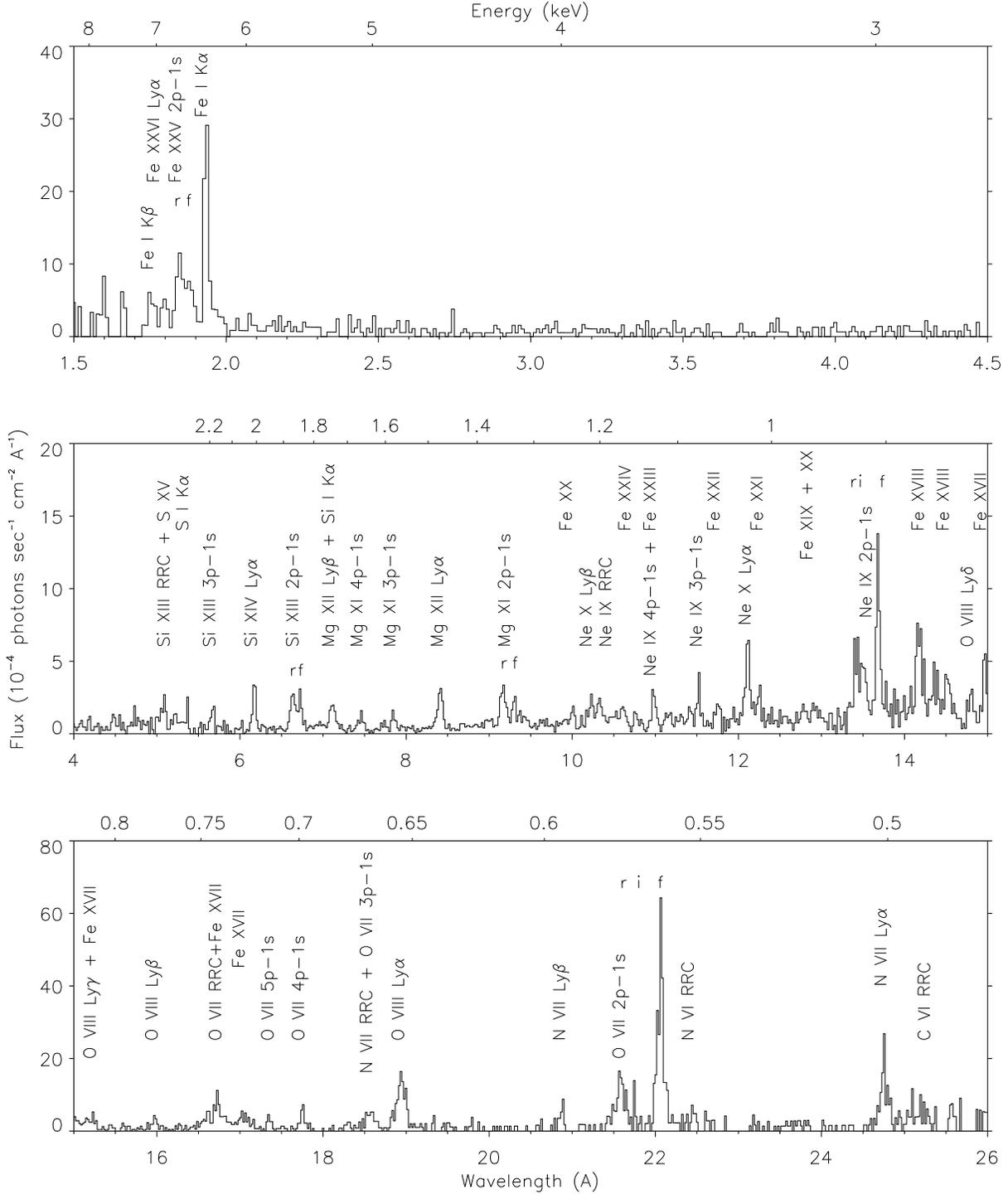} 
\caption{{\it Chandra} HETGS spectra of the central $1.5\arcsec$ radius 
nuclear region. The top panel is from the HEG (0.005 \AA\ bins); 
bottom two are from MEG (0.01 \AA\ bins). The strong forbidden emission lines (f) and 
narrow recombination continuum features (RRC) indicate recombination following photoionization 
by the hidden nucleus. Fe I, Si I, and S I K$\alpha$ emission lines 
indicate fluorescence from a low-ionization region, which may be identified 
with the obscuring molecular torus.}
\label{fig1}
\end{figure*}

\begin{figure*}
\centering
\includegraphics[width=17cm]{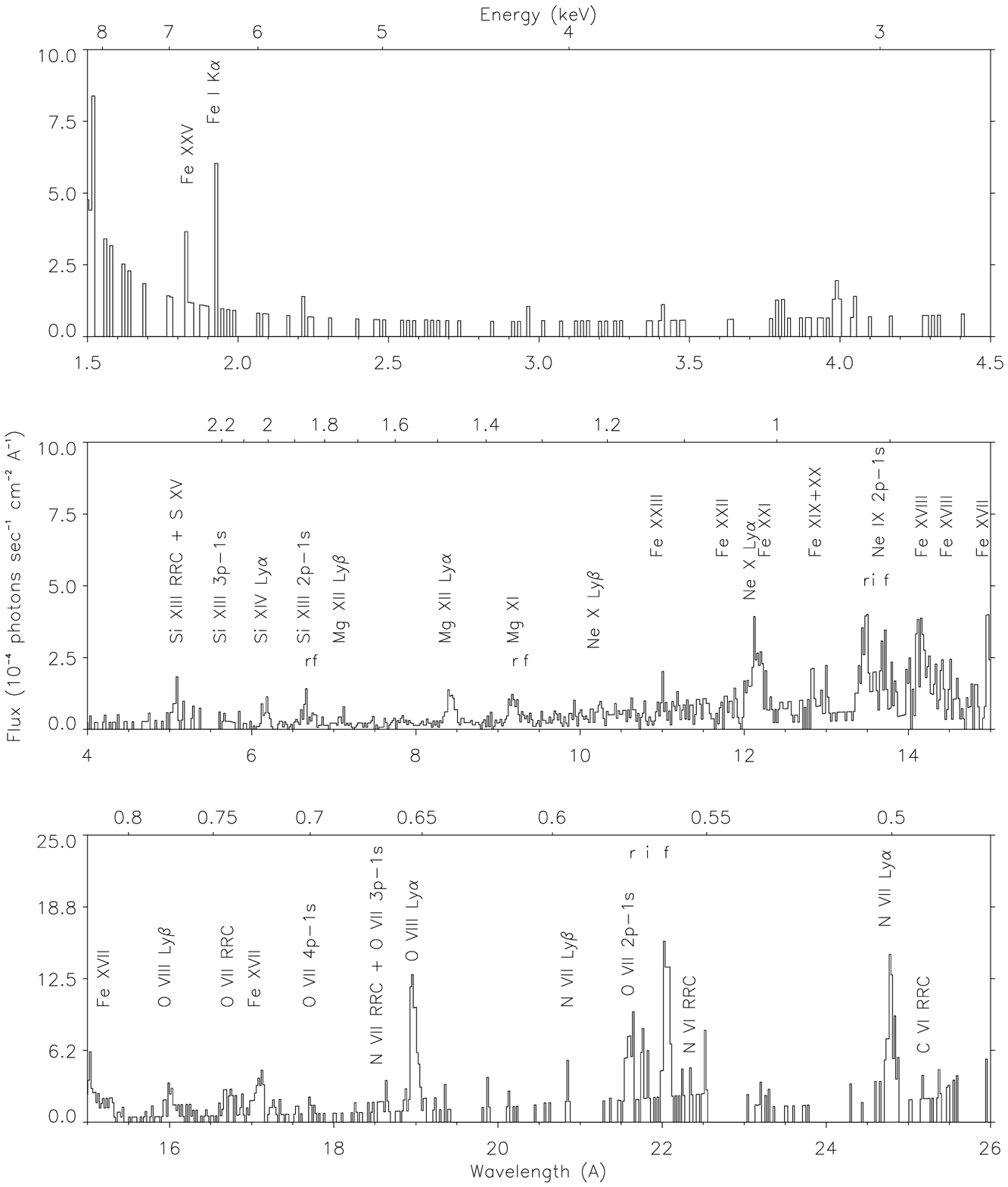} 
\caption{{\it Chandra} HETGS spectra of the NE cloud, centered $3.1 \arcsec$
NE of the nucleus. Binning is the same as in Fig. 1. The forbidden lines (f) and RRCs are
weaker than in the nuclear region (Fig. 1), due to lower column density. The 
H-like and He-like resonance lines are relatively stronger from photoexcitation. 
The Fe L emission lines are stronger, indicating a greater Fe abundance. Fe K$\alpha$ emission 
is weaker because there is relatively little neutral matter in this region.}
\label{fig2}
\end{figure*}

Events were filtered by grade (eliminating grades 1, 5 and 7) and streak 
events were removed from chip S4. First-order HEG and MEG spectra of the 
nuclear region (Fig. ~\ref{fig1}) and off-nuclear region (Fig. ~\ref{fig2}) 
were extracted using CIAO 2.1. Extraction windows in the cross-dispersion 
direction are $3.2 \arcsec$ and $3.0 \arcsec$, respectively. The nuclear 
extraction window is centered on the peak X-ray emission. The off-nuclear 
extraction window is centered on a cloud $3.1\arcsec$ NE of the nucleus (NE 
cloud). The two spectral extraction regions are adjacent and do not overlap. 
Spectra were divided by grating effective area curves generated from the 
aspect histogram and instrument effective area. 

The ACIS zeroth order image provides useful information about the spatial and 
spectral distribution of X-ray emission. The central $3\times3$
pixels have a count rate of 0.31 counts per 3.2 sec frame, which indicates
moderately heavy pileup. This precludes accurate spectral analysis of the 
X-ray peak, but should not greatly affect the surrounding regions. We have 
split the zeroth order image into 6 energy bands: 0.4-0.6, 0.6-0.8, 0.8-1.3, 
1.3-3, 3-6, and 6-8 keV (Fig. ~\ref{fig3}). These bands correspond roughly to 
the O {\sc vii}, O {\sc viii}, Ne \& Fe L, Mg {\sc xi}-S {\sc xv}, hard continuum, 
and Fe K line series, respectively. The individual images are 
$30\arcsec=2.2 h_\mathrm{75}^{-1}$ kpc on a side. We adaptively smoothed the 
broad-band images using the CIAO program CSMOOTH and combined them into two 
3-color images (Figs. ~\ref{fig4}, ~\ref{fig5}), which are $20\arcsec=1.5 
h_\mathrm{75}^{-1}$ kpc on a side. The high-energy image (Fig. ~\ref{fig4}) shows the
distribution of Fe K$\alpha$, scattered continuum, and high-ionization line 
emission. The low-energy image (Fig. ~\ref{fig5}) shows the distributions of 
lower ionization states and Fe L emission.

\section{Nuclear Spectrum}

The nuclear spectrum of \object{NGC 1068} (Fig. ~\ref{fig1}) consists of 
emission from H-like and He-like ions of C, N, O, Ne, Mg, Al, Si, S, and Fe, and
Fe {\sc xvii-xxiv}. Lines were fitted with Gaussians to determine their rest 
wavelengths and fluxes (Table ~\ref{tab1}, ~\ref{tab2}). Positive IDs were made for lines 
within 0.04 \AA\ of the predicted wavelengths for abundant ions. 

We fit the brightest unblended emission lines from the MEG spectrum of the
nuclear region with Gaussian profiles to determine their redshifts and Doppler
$b$ parameters (Table ~\ref{tab3}). The profiles are convolved with a Gaussian spatial
profile, with FWHM$=0.7\arcsec$. We find that all of the lines are blue-shifted,
with $<v>=-410\pm 80$ km s$^{-1}$. This is consistent 
with the {\it HST} [O {\sc iii}] spectrum of the nuclear region \citep{ck00}, 
and shows that the X-ray emitting plasma is flowing out along with the optical
NLR clouds.  All of the H-like and He-like X-ray lines from the nuclear region
are resolved. We find a range of line widths, from $b=690$ km s$^{-1}$ for O 
{\sc vii} to $b=1600$ km s$^{-1}$ for Mg {\sc xii} (Table ~\ref{tab3}). These 
widths are significantly larger than the average optical line widths, but 
consistent with the most extreme  [O {\sc iii}] velocity widths found in 
optical NLR clouds \citep{cdg01}.

Photoionization is indicated by the strength of the forbidden lines and 
prominent narrow radiative recombination continua (RRCs). The ratios $r/f$ of 
resonance to forbidden lines from the He-like ions (e.g. Mg {\sc xi} 2p-1s/f) 
are intermediate  between those expected for pure recombination and 
collisional excitation \citep{pd00}. The nuclear spectrum of \object{NGC 1068}
is very similar in this respect to the spectrum of \object{NGC 4151} 
~\citep{o00}. However, a large value of $r/f$ doesn't by itself indicate a 
collisionally ionized component to the spectrum. As demonstrated with 
{\it XMM} RGS \citep{ksb02}, photoexcitation enhances the high-order resonance
transitions as well as the $n=2-1$ (r) line. The strong high order transitions
in the {\it Chandra} HETGS spectrum of \object{NGC 1068} (Fig. ~\ref{fig1},
Table ~\ref{tab1}, ~\ref{tab2}) confirm that photoexcitation, not collisional excitation is 
responsible for the strong  $n=2-1$ (r) lines of Mg {\sc xi} and Si {\sc xiii}.

Fluorescence is also an important contributor to the nuclear spectrum.  We see
strong, narrow Fe {\sc i} K$\alpha$ and detect Fe {\sc i} K$\beta$, 
S {\sc i} K$\alpha$, and Si {\sc i} K$\alpha$ lines. Fe {\sc i} K$\alpha$ is 
observed at a rest wavelength of $\lambda=1.934\pm0.02$ \AA\ , and the core
is unresolved with a width of $\sigma=4.7^{+0.6}_{-2.2}\times 10^{-3}$ \AA\ . If 
the line is purely from Fe {\sc i}, it is significantly blue-shifted, with 
$v=-500\pm 300$ km s$^{-1}$  and its width corresponds to a turbulent velocity
of $b<1100$ km s$^{-1}$ (90\% confidence). Alternatively, the line may be a 
blend of Fe {\sc i-xviii} \citep{dbo95}, which could cause an apparent 
blue-shift. The S {\sc i} K$\alpha$ line is also blue-shifted, with 
$v=-390\pm 170$ km s$^{-1}$. Fe {\sc i} K$\alpha$ has a weak red Compton shoulder, 
as observed by \cite{ifm97}. The ratio of Fe K$\beta$ to Fe K$\alpha$ is 
$0.17\pm 0.06$, consistent with the predicted ratio ($\sim 0.13$) of fluorescence 
yields \citep{km93}. 

The hard nuclear continuum emission (2.5-6.2 keV) is roughly fit by a power 
law with $\Gamma=1.2 \pm 0.3$ and $F_\mathrm{2-10}=2.8 \times 10^{-12}$ erg cm$^{-2}$ 
s$^{-1}$ ($L_\mathrm{2-10}=7.7 \times 10^{40}$ erg s$^{-1}$), using a (fixed) Galactic 
column of $N_\mathrm{H}=3.5\times 10^{20}$ cm$^{-2}$. This is harder than the direct 
spectrum ($\Gamma=1.8\pm0.2$) from a typical Seyfert 1 galaxy \citep{r02}, and 
indicates neutral Compton reflection, consistent with the strong 
Fe {\sc i} K$\alpha$ fluorescence line \citep{mbf96}. The reflection spectrum 
is analyzed in more detail below.

 \begin{table}[t]
\caption[]{NGC 1068 emission line fluxes ($10^{-5}$ ph s$^{-1}$ cm$^{-2}$) from 
the nuclear region. Predicted wavelengths $\lambda_\mathrm{p}$(\AA) are from 
\cite{gm65}, \cite{js85}, \cite{d88}, \cite{bb98}, and \cite{bb02}. Observed 
wavelength is divided by $1+z_\mathrm{gal}$.}
\label{tab1}
\[
\begin{array}{p{0.13\linewidth}p{0.13\linewidth}p{0.10\linewidth}p{0.10\linewidth}p{0.10\linewidth}p{0.29\linewidth}}
\hline
\noalign{\smallskip}

$\lambda_\mathrm{p}$ &$\lambda_\mathrm{obs}$ &$\pm$ &Flux &$\pm$ &ID \\
\hline    
1.757  &1.755  &0.005 &1.2 &0.5 & Fe {\sc i} K$\beta$   \\
1.780  &1.797  &0.007 &0.9 &0.6 & Fe {\sc xxvi} Ly$\alpha$ \\
1.850  &1.843  &0.005 &1.4 &0.6 & Fe {\sc xxv} 2p-1s (r)$^\mathrm{a}$\\
1.858  &1.854  &0.012 &1.2 &0.7 & Fe {\sc xxv} i $^\mathrm{a}$ \\
1.868  &1.881  &0.005 &2.1 &0.6 & Fe {\sc xxv} f $^\mathrm{a}$  \\
1.9373 &1.9357 &0.0010 &5.1 &0.6 & Fe {\sc i} K$\alpha$ \\
5.039  &5.045  &0.009 &0.6 &0.2 & S {\sc xv} 2p-1s (r) $^\mathrm{a}$ \\
5.086, 5.101 &5.099 &0.005 &0.5 &0.2 & Si {\sc xiii} RRC + S {\sc xv} f \\
5.374, 5.405 &5.367 &0.003 &0.4 &0.2 & S {\sc i} K$\alpha$ + Si {\sc xiii} 4p-1s \\
5.681  &5.676  &0.004 &0.3 &0.1 & Si {\sc xiii} 3p-1s  \\
6.182  &6.181  &0.002 &1.0 &0.1 & Si {\sc xiv} Ly$\alpha$  \\
6.648  &6.636  &0.002 &1.2 &0.2 & Si {\sc xiii} 2p-1s (r) $^\mathrm{a}$\\
6.69   &6.676  &0.004 &0.6 &0.1 & Si {\sc xiii} i $^\mathrm{a}$ \\
6.740  &6.728  &0.002 &1.1 &0.1 & Si {\sc xiii} f $^\mathrm{a}$ \\
7.106  &7.096  &0.006 &0.6 &0.1 & Mg {\sc xii} Ly$\beta$ $^\mathrm{a}$\\
7.130  &7.128  &0.004 &0.6 &0.1 & Si {\sc i} K$\alpha$  $^\mathrm{a}$\\
7.310  &7.318  &0.005 &0.04 &0.04  & Mg {\sc xi} 5p-1s \\
7.473  &7.480  &0.015 &0.2 &0.1 & Mg {\sc xi} 4p-1s \\
7.757  &7.746  &0.003 &0.20 &0.07 & Al {\sc xii} 2p-1s \\
7.850  &7.839  &0.005 &0.5 &0.1 & Mg {\sc xi} 3p-1s \\
8.421  &8.397  &0.005 &2.8 &0.3 & Mg {\sc xii} Ly$\alpha$ \\
9.169  &9.159  &0.004 &1.7 &0.2 & Mg {\sc xi} 2p-1s (r) $^\mathrm{a}$\\
9.23   &9.210  &0.008 &0.7 &0.2 & Mg {\sc xi} i$^\mathrm{a}$ \\
9.314  &9.297  &0.004 &1.2 &0.2 & Mg {\sc xi} f$^\mathrm{a}$ \\
10.007 &10.006 &0.007 &0.5 &0.1 & Fe {\sc xx} \\
\noalign{\smallskip} 
\hline 
\end{array}
\]
\begin{list}{}{}
\item[$^{\mathrm{a}}$] De-blended
\end{list}
\end{table}

\begin{table}
\caption[]{Continuation of Table 1.}
\label{tab2}
\[
\begin{array}{p{0.13\linewidth}p{0.13\linewidth}p{0.10\linewidth}p{0.10\linewidth}p{0.10\linewidth}p{0.29\linewidth}}
\hline
\noalign{\smallskip}

$\lambda_\mathrm{p}$ &$\lambda_\mathrm{obs}$ &$\pm$ &Flux &$\pm$ &ID \\
\hline  
10.239 &10.228 &0.005 &1.3 &0.2 & Ne {\sc x} Ly$\beta$ $^\mathrm{a}$ \\
11.000, &10.984 &0.004 &1.3 &0.2 & Ne {\sc ix} 4p-1s \\
10.981  &       &      &    &    & + Fe {\sc xxiii} 3p-2s\\
11.547, &11.520 &0.004 &1.2 &0.2 & Ne {\sc ix} 3p-1s \\
11.490  &       &      &    &    & + Fe {\sc xxii} 3p-2p\\
12.134 &12.114 &0.004 &3.5 &0.4 & Ne {\sc x} Ly$\alpha$ \\
12.284 &12.257 &0.004 &0.9 &0.2 & Fe {\sc xxi} 3d-2p \\
13.447 &13.430 &0.006 &2.2 &0.6 & Ne {\sc ix} 2p-1s (r) $^\mathrm{a}$ \\
13.518, 13.55 &13.520 &0.006 &2.8 &0.5 & Fe {\sc xix} 3d-2p + Ne {\sc ix} i\\
13.698 &13.680 &0.003 &6.0 &0.6 & Ne {\sc ix} f \\
13.795 &13.777 &0.005 &0.8 &0.3 & Fe {\sc xix} 3d-2p \\
14.034 &14.06  &0.05  &0.7 &0.6 & Fe {\sc xix} 3d-2p \\
14.208 &14.178 &0.008 &4.4 &0.7 & Fe {\sc xviii} 3d-2p \\
14.373 &14.36  &0.02  &1.4 &0.6 & Fe {\sc xviii} 3d-2p \\
14.534 &14.51  &0.02  &1.9 &0.7 & Fe {\sc xviii} 3d-2p \\
14.961 &14.963 &0.005 &1.6 &0.5 & Fe {\sc xix} 3s-2p \\
15.014 &15.024 &0.011 &1.4 &0.5 & Fe {\sc xvii} 3d-2p \\
15.079 &15.053 &0.012 &0.5 &0.5 & Fe {\sc xix} 3s-2p \\
15.176 &15.160 &0.011 &1.2 &0.5 & O {\sc viii} Ly$\gamma$ $^\mathrm{a}$ \\
15.261 &15.232 &0.002 &1.3 &0.3 & Fe {\sc xvii} 3d-2p \\
15.453 &15.447 &0.009 &0.3 &0.2 & Fe {\sc xvii} 3d-2p \\ 
16.006 &15.975 &0.005 &1.0 &0.4 & O {\sc viii} Ly$\beta$  \\
17.051 &17.035 &0.017 &3.4 &1.0 & Fe {\sc xvii} 3s-2p \\
17.200 &17.161 &0.012 &0.6 &0.4 & O {\sc vii} 6p-1s\\
17.396 &17.376 &0.006 &0.3 &0.4 & O {\sc vii} 5p-1s \\
17.768 &17.758 &0.004 &2.3 &0.7 & O {\sc vii} 4p-1s\\
18.629 &18.60  &0.02  &1.4 &1.2 & O {\sc vii} 3p-1s + N {\sc vii} RRC$^\mathrm{a}$ \\
18.969 &18.947 &0.014 &7.7 &1.9 & O {\sc viii} Ly$\alpha$ \\
20.910 &20.883 &0.008 &2.6 &1.2 & N {\sc vii} Ly$\beta $  \\
21.602 &21.589 &0.007 &8.5 &2.1 & O {\sc vii} 2p-1s (r) \\
21.804 &21.747 &0.005 &1.8 &1.3 & O {\sc vii} i $^\mathrm{a}$ \\
22.097 &22.072 &0.002 &20.4 &3.5 & O {\sc vii} f \\
24.781 &24.741 &0.008 &8.0 &2.1 & N {\sc vii} Ly$\alpha$ \\
\noalign{\smallskip} 
\hline 
\end{array}
\]
\begin{list}{}{}
\item[$^{\mathrm{a}}$] De-blended
\end{list}
\end{table}

\begin{table}
\caption[]{NGC 1068 Nuclear Region. Ion velocity $v$ and Doppler parameter $b$ 
(km s$^{-1}$) are from a Gaussian fit. Electron temperature $T_\mathrm{e}(10^4$ K) is from an RRC fit.
Predicted temperature $T_\mathrm{e,p}$ is from an XSTAR photoionization model.}
\label{tab3}
\[
\begin{array}{p{0.13\linewidth}p{0.11\linewidth}p{0.18\linewidth}p{0.18\linewidth}p{0.14\linewidth}p{0.14\linewidth}}
\hline
\noalign{\smallskip}
Ion & Line & $v$  &  $b$ & $T_\mathrm{e}$ & $T_\mathrm{e,p}$\\
\noalign{\smallskip}
\hline 
\noalign{\smallskip}       
{\rm C} {\sc vi} &... &...                &...                   & ... &{\it 2.8} \\
N {\sc vi}     &... &...                  &...                   & ... &{\it 3.2}\\
N {\sc vii}    & Ly$\alpha$ &-350$^{+300}_{-300}$ & 1040$^{+680}_{-340}$ & 3$^{+\infty}_{-2}$&{\it 5.4}\\
O {\sc vii}    & f & -530$^{+110}_{-110}$ &  600$^{+270}_{-170}$ & 8$^{+3}_{-2}$&{\it 3.2} \\
O {\sc viii}   & Ly$\alpha$ &-410$^{+250}_{-250}$ & 1230$^{+420}_{-270}$ &... &{\it 8.5}\\
Ne {\sc ix}    & f & -370$^{+120}_{-110}$ &  750$^{+250}_{-170}$ & 9$^{+11}_{-3}$&{\it 8.5}\\
Ne {\sc x}     & Ly$\alpha$ &-460$^{+300}_{-280}$ &...                   & ...  &{\it 20}\\
Mg {\sc xi}    & f &-300$^{+320}_{-310}$ & 1380$^{+530}_{-300}$ & 7$^{+20}_{-4}$&{\it 13}\\
Mg {\sc xii}   &Ly$\alpha$ &-440$^{+220}_{-230}$ & 1570$^{+400}_{-280}$ &... &{\it 33}\\
Si {\sc xiii}  & f &-540$^{+170}_{-180}$ &  690$^{+400}_{-190}$ &... &{\it 33} \\
Si {\sc xiv}   &Ly$\alpha$ &-330$^{+180}_{-180}$ &  740$^{+450}_{-210}$ &... &{\it 50} \\
\noalign{\smallskip} 
\hline 
\end{array}
\]
\end{table}

\section{Off-Nuclear Spectrum}

The spectrum of the NE cloud (Fig. ~\ref{fig2}) is quite different from the 
nuclear spectrum (Fig. ~\ref{fig1}). One striking difference is the weakness 
of the Fe {\sc i} K$\alpha$ line. This indicates that most of the fluorescence 
occurs in the nuclear region and not in the extended NLR.  

Recombination emission is also considerably weaker in the NE cloud than in the 
nuclear region. Note the weakness of the O {\sc vii} f and Ne {\sc ix} f forbidden
lines and the O {\sc vii} RRC in Fig. ~\ref{fig2}. This indicates a 
relatively low column density of plasma in the NE cloud, which we
model in detail below. Emission from Fe {\sc xvii}-{\sc xxiv} L lines dominates the 
0.7-1.3 keV band. Strong Fe L lines, (e.g. 3d-2p lines of Fe {\sc xvii}) could be 
mistaken as an indicator of collisional excitation. However, as we show below, these 
lines arise from recombination and photoexcitation, as do the lines of the H-like and He-like 
ions. 

We fit unblended lines from the NE cloud with Gaussians, convolved with a 
Gaussian spatial profile with FWHM$=1.5\arcsec$ (estimated from the 0.7-1.3 
keV image). There is no measurable redshift, and the mean velocity is 
$<v>=0\pm 330$ km s$^{-1}$ with respect to the host galaxy rest frame. This is 
consistent with the low (red-shifted) velocities measured for optical 
[O{\sc iii}] emitting clouds in this region by \cite{ck00}. They interpret 
this as evidence for deceleration by the ambient medium in the host galaxy, 
starting at a distance of $1.7\arcsec$ from the nucleus. Alternatively, this 
cloud may mark the shock front where the radio lobe impacts the ISM of the 
host galaxy \citep{cam97, pfbw97,cdg01}. Three emission lines are resolved 
(O {\sc viii} Ly$\alpha$, O {\sc vii} f, and N {\sc vii} Ly$\alpha$), with 
$<b>=810\pm 30$ km s$^{-1}$, similar to values in the nuclear region. A fourth 
line, Mg {\sc xii} Ly$\alpha$, has a rather large Doppler parameter of 
$b=2900\pm 500$ km s$^{-1}$, but may be blended with Fe {\sc xxiii}.

\section{Broad Band Images}

\subsection{Comparison to H$\alpha$}

To put our X-ray images in the context of the vast lore of optical emission
line studies of \object{NGC 1068}, we have combined them with an archival 
{\it HST} WFPC2 H$\alpha$ image (Fig. ~\ref{fig6}). The {\it HST} image was 
taken through the F658N filter for 900 s, and the estimated continuum (F791W 
filter) was subtracted out \citep{cam97, k99a}. Red represents H$\alpha$ + 
N {\sc ii}, green soft X-rays (0.4-0.7 keV), and blue the 0.7-1.3 keV band. The
peak of the X-ray image was aligned to match cloud B \citep{ev91} in the 
H$\alpha$ image. This is the closest cloud to the hidden nucleus, as 
determined by UV polarimetry \citep{k99b}. Note that this is slightly 
different from \cite{yws01} who align the Chandra ACIS images with the radio 
core. However, this makes little difference at Chandra resolution 
($0.5\arcsec$).

Chandra ACIS images \citep{yws01} show a strong correspondence between the 
optical [O III] and X-ray emission in the NLR. We find a weaker correspondence
between the H$\alpha$ and X-ray emission. In fact, it appears that the
highly ionized X-ray NLR is surrounded by a cocoon of H$\alpha$ filaments.
The tip of this cocoon, $5\arcsec$ NE of the nucleus, corresponds to the 
tip of the radio lobe \citep{wu87,cam97,yws01}. The bright X-ray cloud 
$3\arcsec$ NE of the nucleus (NE cloud) lies in the back-flow region of the 
radio lobe, but does not correspond to any individual cloud in the H$\alpha$ 
or [O {\sc iii}] images. The NE cloud dominates the off-nuclear HETGS spectrum
(Fig. ~\ref{fig2}). 

The spiral structure of the H$\alpha$ emission appears to be associated with 
the central few arc seconds of the starburst disk in the host galaxy 
\citep{cam97}. The NE cone of the X-ray NLR is seen projected in front of
the disk, while the SW cone is hidden behind it. The SE spiral arm in the 
0.8-1.3 keV X-ray image (Fig. ~\ref{fig3}) matches the arm seen in H$\alpha$, 
consistent with energetic starburst activity. The starburst disk also extends 
out to much larger radius, which is much better seen in the ACIS image of 
\cite{yws01}. 

\subsection{Hard X-ray Image}

\begin{figure}
\resizebox{\hsize}{!}{\includegraphics{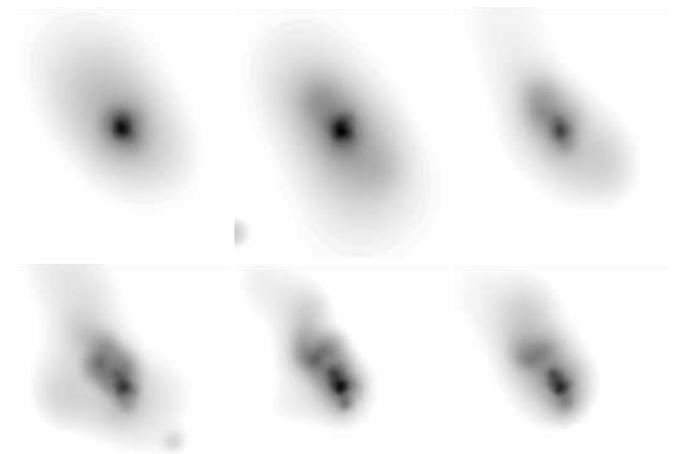}}
\caption{Six broad-band images filtered from the HETGS zeroth order image and
adaptively smoothed.
First row: 6-8, 3-6, 1.3-3 keV. Second row: 0.8-1.3, 0.6-0.8, 0.4-0.6 keV. Each 
image is $30\arcsec=2.2 h_\mathrm{75}^{-1}$kpc on a side, 0.49\arcsec/pixel. Note the
change in structure with energy. Fe K$\alpha$ emission is concentrated in
the nuclear region in the 6-8 keV image. Subtle differences between the 
various soft X-ray bands indicate different spatial distributions for various 
ionization states. Starburst emission from the host galaxy disk is most apparent 
east of the nucleus in the 0.8-1.3 keV image.}
\label{fig3}
\end{figure}

\begin{figure}
\resizebox{\hsize}{!}{\includegraphics{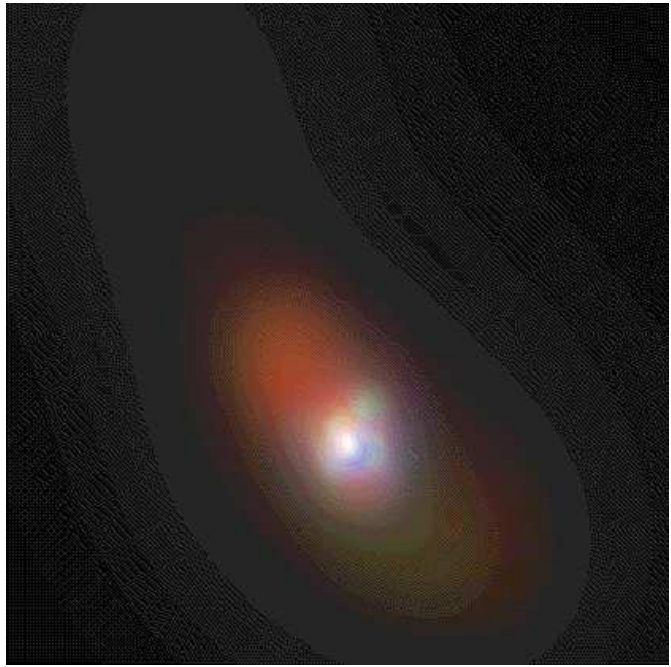}}
\caption{Three-color high energy X-ray image ($E=$1.3-3, 3-6, 6-8 keV)$=$
(r,g,b). The image is $20\arcsec=1.5 h_\mathrm{75}^{-1}$ kpc on a side. The brightest
point-like source of high-energy emission may be the inner wall of the
molecular torus, reflecting X-rays from the hidden nucleus. The hard emission 
surrounding the hard X-ray peak may delineate reflection from the outer 
reaches of the obscuring torus. Extended emission to the NE and SW corresponds
to the ionization cone. The SW cone is seen through and partly obscured by the disk of 
the host galaxy.}
\label{fig4}
\end{figure}

The image in the 6-8 keV band (Fig. ~\ref{fig3}) is dominated by Fe {\sc i} 
K$\alpha$ emission, though there is also significant contribution from Fe 
{\sc xxv} (see Fig. ~\ref{fig1}). This emission is spatially resolved, as 
first shown by \cite{yws01}. There is a weak ($\sim 10\arcsec$) linear 
extension in the 6-8 keV band (Fig. ~\ref{fig3}) corresponding to the extended
ionization cones, seen on both sides of the nucleus. The peak of Fe K emission
is close to the hidden nucleus (Fig. ~\ref{fig4}), and may correspond to the 
illuminated inner wall of the torus.

The X-ray peak is surrounded by extended ($2.7\arcsec\times 2.2\arcsec$) emission 
in the 3-8 keV band (Fig. ~\ref{fig4}). We attribute this to the outer edge of an 
irregular molecular torus seen in scattered hard X-rays. This feature corresponds to a 
region of high optical extinction in {\it HST} images and lies at the same 
radius ($\sim 1\arcsec$) as the molecular ring seen in CO emission 
\citep{s00}. The location and size of the hard X-ray ring is also consistent 
with shadowing of the far (SW) scattering cone observed in the near-IR and 
attributed to the torus ~\citep{yph96}.

Emission from the ionization cones is also seen in the 1.3-3 keV and 3-6 keV 
bands (Fig. ~\ref{fig3}, green and red in Fig. ~\ref{fig4}). The 1.3-3 keV 
image is dominated by emission from highly ionized Mg, Si, and S, while the 
3-6 keV emission is mostly scattered continuum. The 3-6 keV continuum photons 
(green) penetrate the ISM of the host galaxy disk, revealing the far-side 
ionization cone which is mostly invisible at lower energies. This is 
consistent with the $N_\mathrm{H}$ map produced by \cite{yws01}, 
and near-IR imaging which first revealed the SW cone \citep{yph96,pyh97}. The 
far cone is mostly obscured at optical-UV and soft X-ray energies, though the 
cloud $1.5\arcsec$ SW of the nucleus (Fig. ~\ref{fig5}) peeks through. 

\subsection{Soft X-ray Image}

\begin{figure}
\resizebox{\hsize}{!}{\includegraphics{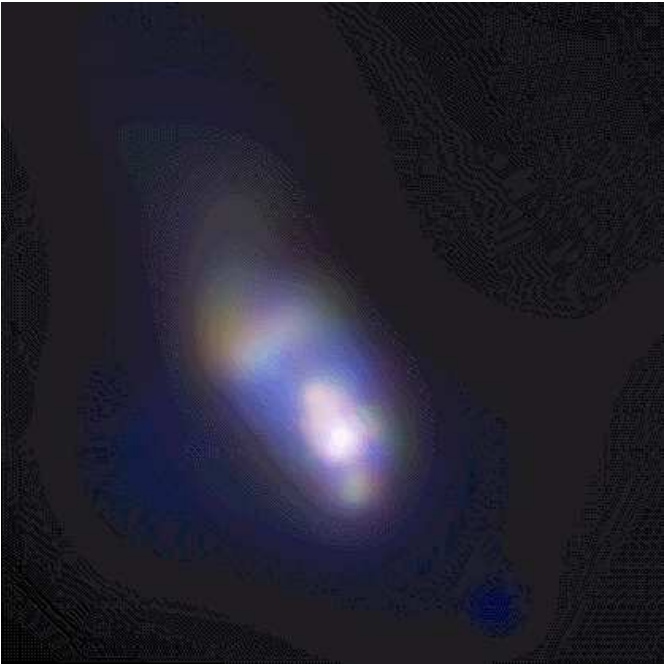}} 
\caption{Three-color low energy X-ray image ($E=$0.4-0.6, 0.6-0.8, 0.8-1.3 
keV)$=$(r,g,b), showing the distribution of line emission, which peaks near the
nucleus. The color of the X-ray NLR ranges from blue to red, corresponding to 
high through lower ionization states. Same scale as Fig. ~\ref{fig4}.}
\label{fig5}
\end{figure}

The X-ray color varies significantly across the low-energy image 
(Fig. ~\ref{fig5}), owing to variations in emission line strengths. These 
variations can be attributed to ionization or abundance variations. The region 
extending $1.5\arcsec$ NE of the nucleus is pink, owing to relatively 
weak 0.8-1.3 keV emission, and suggesting a relatively low ionization 
parameter. This corresponds closely to the strongest [O {\sc iii}] 
emission (arrowhead shaped) from the nuclear region in {\it HST} images 
\citep{ev91,mcs94}. 

\begin{figure}
\resizebox{\hsize}{!}{\includegraphics{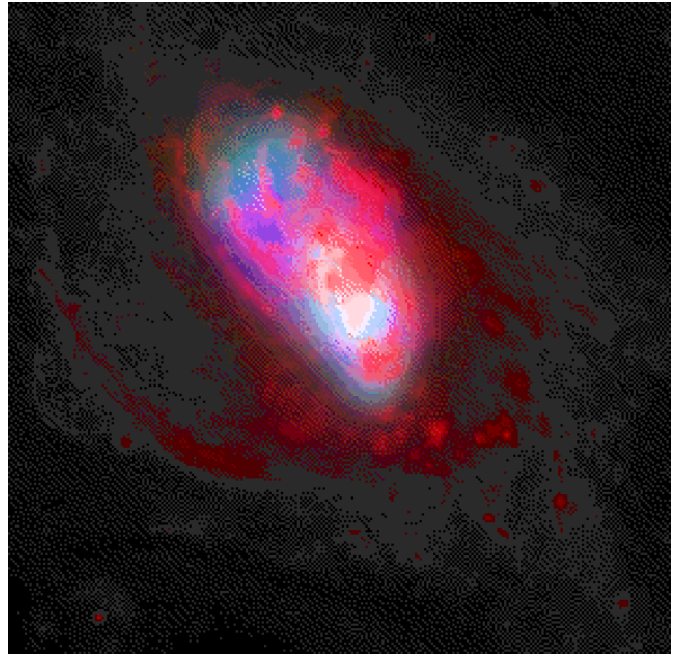}} 
\caption{Composite optical H$\alpha$ (red), X-ray 0.4-0.7 keV (green), 
and X-ray 0.7-1.3 keV (blue) image. The H$\alpha$ image is from {\it HST}, 
while the lower resolution X-ray images are from {\it Chandra}. 
Higher ionization X-ray emission line regions are enveloped by a cocoon
of lower ionization H {\sc ii} filaments. Regions of strong H$\alpha$ emission
in the NLR correspond to the softer X-ray emission regions in Fig. ~\ref{fig5}.
H {\sc ii} regions outside the photoionization cone delineate the
spiral arms in the starburst disk of the host galaxy.}
\label{fig6}
\end{figure}

The NE cloud changes in color from blue to green to red along the radius 
vector from the nucleus (Fig. ~\ref{fig5}), perhaps indicating a change in the 
ionization level. Red regions in Fig. ~\ref{fig5} correspond to regions of 
strong H$\alpha$ emission (Fig. ~\ref{fig6}), which is indeed consistent with 
lower ionization. The NLR clouds $1.5\arcsec$ SW, and $1\arcsec$ W of the 
nucleus (Fig. ~\ref{fig5}) show a similar effect. The lowest ionization (red) 
regions in the NE, SW, and W clouds may indicate the intersection of the 
ionization cone with high density clouds in the disk of the host galaxy (See 
Fig. 6 of \cite{cdg01}). 

The blue gap between the nuclear region and the NE cloud is filled with strong
1 keV emission. There is also strong H$\alpha$ emission in this region 
(Fig. ~\ref{fig6}). There are a few possible interpretations. 1) This could be a 
region of low density and high ionization which emits strongly in Ne {\sc x} and
Fe L  lines. If so, the H$\alpha$ filaments which have lower ionization 
may just be seen in projection. 2) The emission could come from collisionally ionized 
plasma in the starburst disk. The H$\alpha$ emission could then be from star forming regions.  
3) This region could be shock-heated by the NLR outflow or radio jet. The radio jet appears
to be deflected in this region, which may lend support to this interpretation
\citep{yws01}.

\section{Discussion}

\subsection{Photoionization Model}

As seen in other Seyfert 2s \citep{skp00,snk01}, photoionization, recombination, 
and photoexcitation appear to dominate the spectrum of NGC 1068. The main indicators are: 
(1) narrow recombination continua of N {\sc vii}, O {\sc vii}, and Ne {\sc ix}, (2) strong 
forbidden lines of O {\sc vii} and Ne {\sc ix} (3) strong np-1s (n$>$2) lines from O {\sc vii}, 
O {\sc viii}, Ne {\sc ix}, Mg {\sc xi}, and Si {\sc xiii}, (4) weak Fe L emission
(relative to collisionally ionized plasmas). 
We therefore start with a pure photoionization model. In a subsequent section we assess 
the possibility of additional emission from a collisionally ionized component.

We model the nuclear and off-nuclear spectra of \object{NGC 1068} using the 
method of \cite{ksb02}. The geometry is a collection of clouds photoionized
by an external X-ray source. Each cloud is characterized by its radial 
velocity, turbulent velocity, temperature, column density, and source covering 
fraction. The source is characterized by its 2-10 keV X-ray luminosity 
$L_\mathrm{x}$ and power-law spectral index, which we fix at $\Gamma=1.8$, a 
typical value for Seyferts. The velocity and turbulent velocity determine the 
line redshifts and widths.  The electron temperature $T_\mathrm{e}$ controls 
the width of the radiative recombination continuum. The column density 
$N_\mathrm{i}$ and Doppler $b$ parameter determine the relative line 
contributions from recombination and photoexcitation. Note that the 
photoexcitation rate saturates at a larger column density for larger values of
$b$. Line intensity is proportional to column density, covering fraction 
$f_\mathrm{c}$, and X-ray luminosity. 

We fit the spectra to derive the flux-weighted mean cloud parameters for each
ion. Spectral fitting is done with IMP, a code developed at UCSB which employs a 
$\chi^2$ minimization algorithm. For H-like and He-like ions, ($n=1-7$, l) level 
populations are computed by solving rate equations including terms from photoionization, 
recombination, and photoexcitation \citep{bks01}. Recombination rates for $n>7$ are
extrapolated using the hydrogenic approximation. The emission line spectrum of 
each ion is computed from the level populations and transition rates. The radiative 
recombination continua (RRCs) are computed using the method of \cite{l99} and the
photoionization cross sections of ~\cite{vfk96}. Atomic data, including energy levels, 
photoionization rates, recombination rates, and Einstein coefficients were calculated 
using the Flexible Atomic Code (FAC) 0.7.6 ~\citep{gu02a}. Some He-like emission line 
wavelengths were replaced by more accurate values from ~\cite{vvf96}. 

We use a more approximate treatment of Fe {\sc xvii}-{\sc xxiv}. Radiative recombination 
(RR) and dielectronic recombination (DR) line power for these ions are computed by 
\cite{gu02b}, using atomic data from FAC. Photoionization and recombination are balanced to 
determine relative abundances of the $i$ and $i+1$ ionic charge states. We add 
photoexcitation from the ground level to $n=2-5$ and include line emission from the 
resulting cascades. Similar to the H-like and He-like ions, the relative fluxes of
photoexcitation and recombination lines depend strongly on the column density and can
therefore be used to constrain it ~\citep{sk02}.

We start our fit by fixing the Doppler $b$ parameters in the nuclear region at the 
values in Table ~\ref{tab3}, and fixing the emission redshift at its mean value 
$<z>=0.0025$. The Ne {\sc x}, S {\sc xv}, and Fe {\sc xvii}-{\sc xxiv} lines are all 
strongly blended. We fix their widths at $b\sim800$ km s$^{-1}$, which appears to give a 
good overall model fit. Because the Fe {\sc xxv} lines are strongly blended and 
the $n>2$ Fe {\sc xxvi} lines are not detected, the column densities of these ions are not
constrained and we do not include them in our fit. We fit Gaussian profiles for the the S {\sc i} 
and Si {\sc i} K$\alpha$ lines, so that they don't skew the results for lines they are 
blended with. A fixed scattered continuum component ($\Gamma=1.8$) is added to the model, 
extrapolated from the fits of the HEG continuum (see below), so that the emission line fluxes 
aren't overestimated.

Next we fit the RRC profiles to determine the electron temperature. (The width of an RRC 
in eV gives a rough estimate of the temperature of recombining electrons.) The 
O {\sc vii} and Ne {\sc ix} RRCs are well measured, with $T_\mathrm{e}=8-9\times10^4$ K 
(Table ~\ref{tab3}). A larger range of temperatures is probably present for other 
ions observed that have RRCs which are too weak or blended to measure. Therefore, 
we fix $T_\mathrm{e}$ for each of these ions at a value which corresponds to an ionization 
parameter of peak emissivity. These temperatures were computed using the XSTAR 
photoionization code (Table ~\ref{tab3}). Finally, we fit for column densities and 
the quantity $f_\mathrm{c}L_\mathrm{x}$ for each ion (Table ~\ref{tab4}).
 
The fitting procedure is similar for the off-nuclear (NE cloud) spectrum. Since not
all emission lines are resolved, we assume $b=800$ km s$^{-1}$, the mean value
for the resolved lines. The model spectrum is convolved with a spatial profile with 
FWHM$=1.5\arcsec$. Because the RRCs are so weak, we can not derive accurate values 
for the electron temperature in the NE cloud. Again, we assume $T_\mathrm{e}$ values 
derived from XSTAR (Table ~\ref{tab3}). 

\subsection{Ionic Column Densities and Abundances}

\begin{figure*}
\centering
\includegraphics[width=17cm]{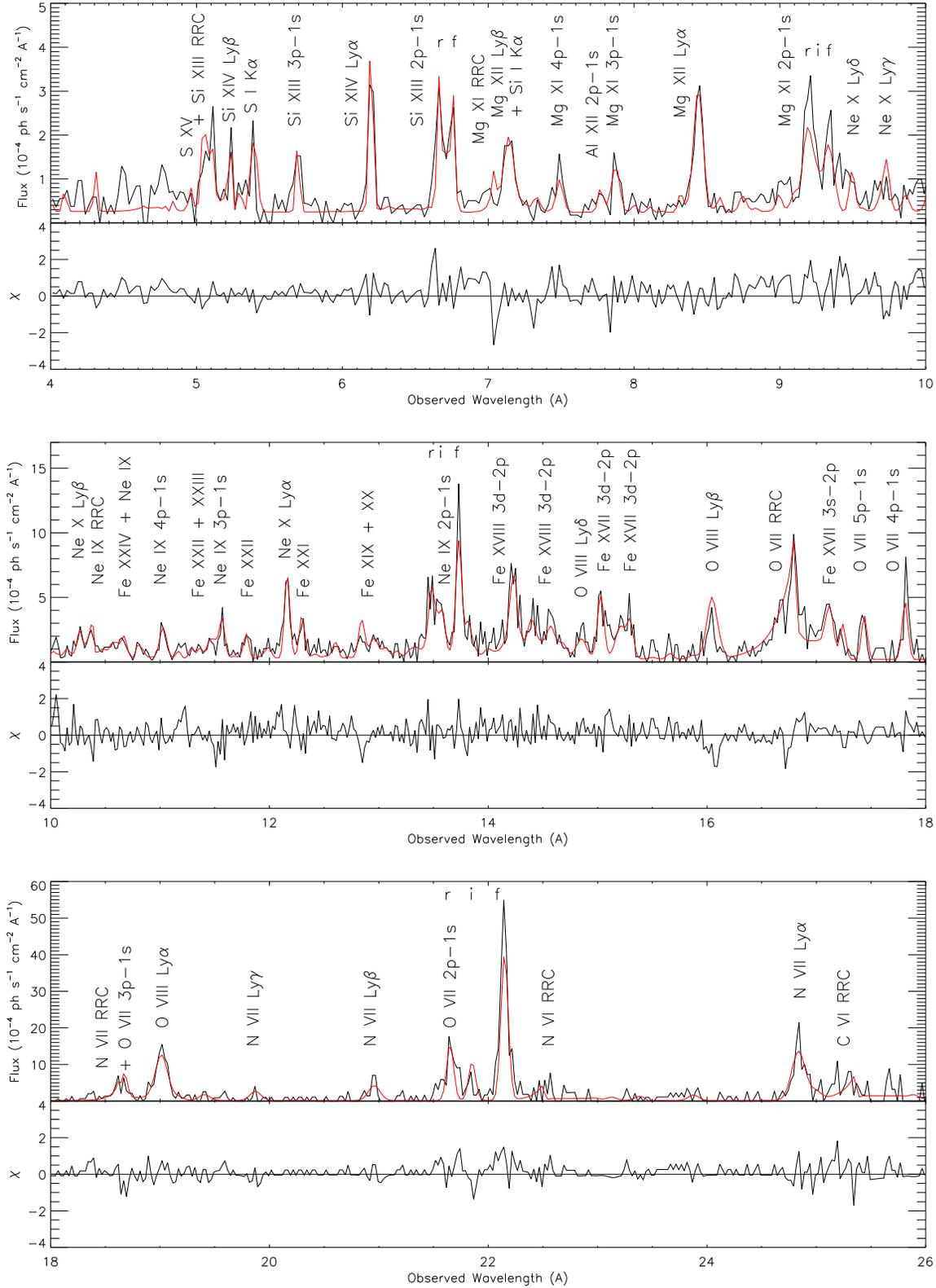}
\caption{Photoionization model of MEG spectrum of nuclear region. 
Ion spectra are fit with IMP, to determine column densities and 
temperatures (see text). The ratios of photoexcitation to recombination
lines are the primary diagnostics for column density. The widths
of the RRCs are a measure of the electron temperature.}
\label{fig7}
\end{figure*}

Fig. ~\ref{fig7} shows our photoionization model fit for the nuclear region. The fit  
is quite good, with small residuals. Notable exceptions are the forbidden (f)
line fluxes of O {\sc vii} and Ne {\sc ix}, which appear to be underestimated by about 
20\%. We conjecture that this may be due to non-Gaussian line shapes. In particular,
the O {\sc vii} lines appear to have underlying broad components which we do not 
model. The opposite problem occurs with O {\sc viii} Ly$\beta$, which is overestimated
because it has a narrower width than O {\sc viii} Ly$\alpha$. This is difficult
to understand, unless O {\sc viii} Ly$\alpha$ is broadened more by scattering or there
are additional emission sources which add to the line. There appears
to be unmodeled excess emission in the 8.8-9.5 \AA\ region, in the vicinity of the
Mg {\sc xi} triplet, which may affect our column density estimate of this ion. 

We find nuclear ionic column densities (Table ~\ref{tab4}) in the range $2\times 10^{16}-6\times 
10^{18}$ cm$^{-2}$, which are comparable to warm absorber column densities observed in 
Seyfert 1 galaxies (e.g., \cite{kml00}). We do not give formal uncertainties for the 
column densities, because it is hard to calculate correlated uncertainties in a $\sim40$ 
parameter fit. However, we do give (90\%) single-parameter uncertainties for 
$f_\mathrm{c} L_\mathrm{x}$. We find $<f_\mathrm{c}L_\mathrm{x}>=1.5 \pm 0.2\times 
10^{41}$ erg  s$^{-1}$ for the nuclear region.

The large $r/f$ ($\ge 1$) values observed for Mg {\sc xi} and Si {\sc xiii} are
nicely explained by the photoionization plus photoexcitation model. The 
abundances and column densities of these ions (Table ~\ref{tab4}) are relatively 
low so photoexcitation dominates over recombination. This is in contrast to the O {\sc vii}
and Ne {\sc ix} ions, which have larger column densities and saturated 2p-1s resonance
lines. There is therefore no need to invoke collisional excitation to explain 
the line ratios. We also reemphasize the fact that the strong 3p-1s, 4p-1s, and 5p-1s
lines require photoexcitation, and would not be present in a collisionally excited plasma.

The column densities we derive for the nuclear region are a factor of a few higher than 
those derived by \cite{ksb02}, who use a smaller radial turbulent velocity of
$\sigma_v=200$ km s$^{-1}$ ($b=280$ km s$^{-1}$) and an additional line of
sight velocity broadening of $\sigma_v=400$ km s$^{-1}$. It seems unlikely
to us that the  turbulent velocity is larger perpendicular to the axis of
the ionization cone than parallel to it. However, it is possible that
bulk motions of clouds may broaden the lines without affecting the resonance
line optical depths in individual clouds. In that case, our measured line
widths give an upper limit to $b$, and our column densities are upper limits.
As pointed out by \cite{sk02}, since the line fluxes are proportional to column density,
fit column densities will be biased toward large values.

\begin{figure*}
\includegraphics[width=17cm]{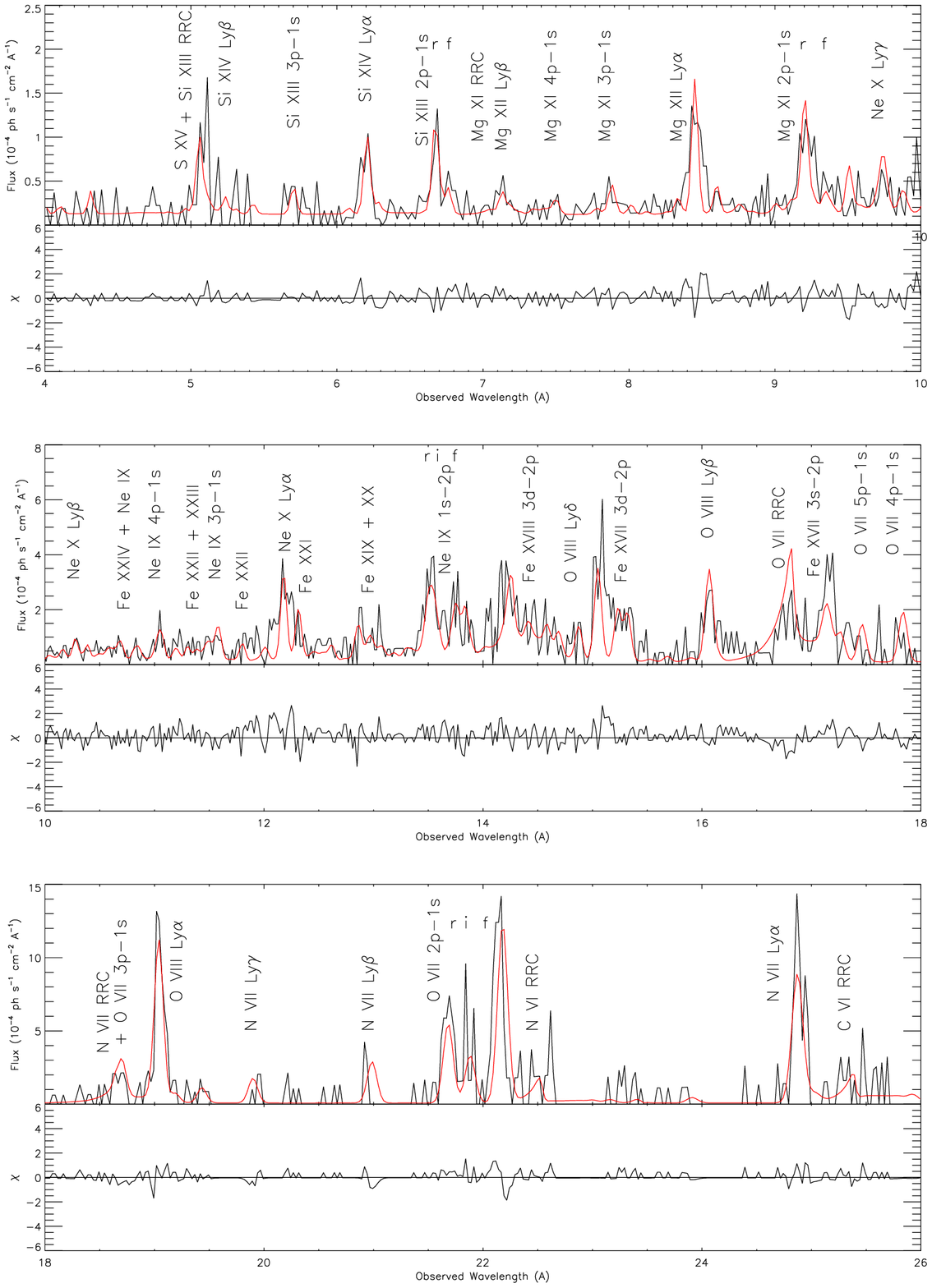}
\caption{Photoionization model of MEG spectrum of the NE cloud. Similar to Fig. 
 ~\ref{fig7}.}
\label{fig8}
\end{figure*}

Fig. ~\ref{fig8} shows our photoionization model fit for the NE cloud. The model does a 
fairly good job of representing the data, except for residuals in the 11.7-12.4 and 15-15.2 
\AA\ regions. The Fe {\sc xxi} emission falls in the former band, precluding an accurate
determination of the column density. The Fe {\sc xvii} column density is underestimated by
the model ($N_\mathrm{i}=4\times 10^{16}$ cm$^{-2}$). This is apparent from comparing the strengths 
of the 3s-2p (17.1 \AA\ ) and 3d-2p (15.0 \AA\ ) lines which indicate a more consistent value of 
$N_\mathrm{i}=2\times10^{17}$ cm$^{-2}$. IMP tried to decrease the unidentified residual at 15 \AA\ by 
strengthening the 3d-2p line with respect to the 3s-2p line, leading to a column density
estimate which is too low. The Al {\sc xii}, S {\sc xv}, and Fe {\sc xxiv} emission are too 
weak to constrain column densities. We also find an unexplained shift in the wavelength of 
the O {\sc vii} f line, though the other lines from this series are consistent with zero 
redshift. 

The covering fraction of the NE cloud is indistinguishable 
from the nuclear region, with $f_\mathrm{c}L_\mathrm{x}=9.0\pm 0.7\times 10^{40}$ erg s$^{-1}$. 
The small dispersion (30-50\%) in $f_\mathrm{c} L_\mathrm{x}$ in both regions indicates that all 
ions have a similar covering fraction, assuming the source is isotropic within the ionization 
cone.
         
We estimate the equivalent hydrogen column $N_\mathrm{H}$ and emission measure
EM$=\int n_\mathrm{e}^2 dV$ for each ion (Table ~\ref{tab4}), assuming solar abundances and 
ion fractions modeled by XSTAR. The mean column densities averaged over all ions are 
$N_\mathrm{H}=5 \pm 1 \times 10^{22}$ cm$^{-2}$ in the nuclear region and 
$N_\mathrm{H}=1.4 \pm 0.3 \times 10^{22}$ cm$^{-2}$ in the NE cloud. The factor of 4 lower mean
column density in the NE cloud accounts for the weaker forbidden lines and RRCs relative to the
nuclear spectrum.  Photoexcitation is relatively stronger because the resonance 
lines are not saturated. The relative column densities of He-like Mg and Si are even 
lower  in the NE cloud, accounting for their very large $r/f$ ratios. Such large ratios are 
impossible to produce by collisional ionization in a low-density plasma. In contrast, relatively 
large Fe ion column densities in the NE cloud account for the strong Fe L-shell emission.

\begin{table}
\caption[]{NGC 1068 relative covering fraction, column density, and emission measure of
the nuclear region (Nuc.) and NE cloud (NE). The quantity $f_\mathrm{c} L_\mathrm{x}$ is normalized 
to the mean for the nuclear region, $<f_\mathrm{c} L_\mathrm{x}>=1.5\times10^{41}$. It therefore 
gives the relative covering fraction, assuming $L_\mathrm{x}$ is the same for all ions. Ionic column 
densities $N_\mathrm{i}$ are given in units of $10^{17}$ cm$^{-2}$ and equivalent hydrogen column 
densities in units of  $10^{22}$ cm$^{-2}$. As explained in the text, IMP underestimates the column 
density of Fe {\sc XVII} in the NE cloud. The value presented in parentheses here gives a better match 
to the spectrum. The emission measure EM is in units of $10^{63}$ cm$^{-3}$.} 
\label{tab4}
\[
\begin{array}{p{0.13\linewidth}p{0.16\linewidth}p{0.06\linewidth}p{0.06\linewidth}p{0.07\linewidth}p{0.16\linewidth}p{0.06\linewidth}p{0.06\linewidth}p{0.07\linewidth}}
\hline
\noalign{\smallskip}
Ion &Nuc. $f_\mathrm{c} L_\mathrm{x}$ & $N_\mathrm{i}$ &$N_\mathrm{H}$ & EM & NE $f_\mathrm{c} 
L_\mathrm{x}$ & $N_\mathrm{i}$ & $N_\mathrm{H}$ & EM \\ 
\hline
\noalign{\smallskip}
N {\sc vii}      &0.8$\pm$0.2& 49  & 8.3  & 16 &0.7$\pm$0.2& 27 & 4.6 & 11\\
O {\sc vii}      &1.7$\pm$0.2& 37  & 0.77 & 20 &0.9$\pm$0.2& 30 & 0.62 & 7.6 \\
O {\sc viii}   &0.51$\pm$0.09& 59  & 1.3  & 2.1&0.40$\pm$0.07& 98& 2.2 & 1.8 \\
Ne {\sc ix}    &0.45$\pm$0.07& 49  & 7.4  & 7.2&0.3$\pm$0.1& 14& 2.1   & 1.4 \\
Ne {\sc x}       &1.1$\pm$0.2& 6.1 & 1.0  &0.76&0.6$\pm$0.2& 5.3& 0.87 & 0.36\\
Mg {\sc xi}    &0.39$\pm$0.07& 23  & 12   & 3.9&0.8$\pm$0.3& 1.2& 0.61 & 0.62\\
Mg {\sc xii}   &0.62$\pm$0.09& 10  & 5.4  & 1.4&0.9$\pm$0.2& 1.7& 0.91 & 0.33\\
Al {\sc xii}     &1.3$\pm$1.0& 0.20& 1.4  & 1.4&...        &...     &... &...\\
Si {\sc xiii}    &0.8$\pm$0.1& 19  & 11   & 3.6&0.8$\pm$0.3& 2.3& 1.3  & 0.42\\
Si {\sc xiv}     &0.7$\pm$0.1& 14  & 8.3  & 1.5&0.6$\pm$0.3& 2.7& 1.6  & 0.22\\
S {\sc xv}       &0.6$\pm$0.4& 20  & 25   & 3.4&...        &...     &... &... \\
Fe {\sc xvii}    &2.5$\pm$0.6& 0.44& 0.30 &0.72&...        &(2) &... &...     \\
Fe {\sc xviii}   &1.5$\pm$0.4& 0.81& 0.66 & 1.0&0.4$\pm$0.1& 1.5& 1.2 & 0.51\\
Fe {\sc xix}     &0.9$\pm$0.3& 1.1 & 2.6  &0.56&0.8$\pm$0.2& 0.98& 2.3 & 0.43\\
Fe {\sc xx}      &0.8$\pm$0.3& 1.3 & 1.0  &0.27&0.5$\pm$0.2& 0.83& 0.66 & 0.11\\
Fe {\sc xxi}     &0.7$\pm$0.3& 2.3 & 3.2  &0.15&...        &...     &...  &... \\
Fe {\sc xxii}    &1.9$\pm$0.7& 0.49& 0.35 &0.08&0.9$\pm$0.6& 0.38& 0.28 & 0.027\\
Fe {\sc xxiii}   &0.8$\pm$0.5& 1.1 & 0.97 &0.09&0.4$\pm$0.4& 0.81& 0.71 & 0.031\\
Fe {\sc xxiv}    &1.0$\pm$0.9& 0.62& 0.89 &0.05&...        &...     &... &... \\

\hline
\end{array}
\]
\end{table}

\begin{table}
\caption[]{NGC 1068 relative elemental abundances.}
\label{tab5}

\[
\begin{array}{p{0.17\linewidth}p{0.17\linewidth}p{0.17\linewidth}p{0.17\linewidth}p{0.17\linewidth}}
\hline
\noalign{\smallskip}
Element &Nuclear Region &NE cloud &Solar \\ 
\hline
\noalign{\smallskip}
N        & $>0.89$ & $>1.40$ & 0.91 \\
O        & 1.74    & 6.63    & 6.92 \\
Ne       & 1.00    & 1.00    & 1.00 \\
Mg       & 0.60    & 0.15    & 0.31 \\
Al       &$>0.004$ &...      & 0.024\\
Si       & 0.60    & 0.26    & 0.29 \\
S        &$>0.36$  &...      & 0.13 \\
Fe       &0.15     &$>0.34$  & 0.38 \\
\hline
\end{array}
\]
\end{table}

Relative elemental abundances (Table ~\ref{tab5}) are estimated by summing the
ionic columns and taking the ratio with respect to neon. These are rough estimates 
since we are comparing clouds with a large range of ionization states, and
ignore possible abundance gradients. Except for oxygen and iron, 
the relative abundance patterns for both the nuclear region and NE cloud are similar 
(within a factor of 2) to solar. Oxygen appears to be under-abundant by a factor of 4.0
in the nuclear region, but is near solar abundance in the NE cloud. Oxygen depletion by 
a factor of 5 was previously inferred ~\citep{m93}, based on the weakness of 
the O {\sc viii} Ly$\alpha$ line. A similar depletion by a factor of 4.0 has been inferred for 
nuclear molecular clouds based on a large HCN/CO intensity ratio ~\citep{sgt94}.
The Fe abundance is depleted by a factor of 2.5 in the nuclear region, but is near solar 
abundance in the NE cloud.

\subsection{Limits on Collisionally Ionized Plasma}

Our observations confirm that line emission from the X-ray NLR in 
\object{NGC 1068} is dominated by warm ($T_\mathrm{e}=8\times 10^4$) photoionized gas. 
However, past observations indicate that shock-heating may also
be important in the NLR. The NE radio lobe terminates at a distance of $\sim 
5\arcsec$ from the nucleus and  has a sharp leading edge, possibly 
corresponding to a shock front \citep{wu87}. There are also indications that 
the NLR clouds decelerate on the $2-4 \arcsec$ scale \citep{ck00}, perhaps by 
plowing through the ISM of the host galaxy. Other studies have shown that 
photoionization by emission from shocks may be important in the optical NLR 
\citep{ds95}. In addition, pressure confinement of the NLR clouds may require 
the presence of a hot component with $T_\mathrm{e}\simeq10^7$ K 
\citep{kmt81,o00}. 

Adding a collisionally ionized (CI) component to the spectrum would increase 
the strength of the 2p-1s ($\alpha$) lines of H-like and He-like ions. The 
contribution to forbidden, inter-combination, and high-order Lyman lines 
would be relatively weaker \citep{ksb02}. Since the line ratios observed in 
\object{NGC 1068} are closely consistent with a photoionized plasma, the 2$\sigma$ 
uncertainties in the $n=2-1$ line fluxes give reasonable upper limits to the 
contribution from CI. (Similar estimates could be made from the residual broad
O {\sc vii} or O {\sc viii} emission.)  We use the emissivities and peak temperatures 
of \cite{mg85}, who assume solar abundances. From the O {\sc vii} 2p-1s line 
(Table ~\ref{tab2}), the 2$\sigma$ upper limit for the emission measure of a 
collisionally ionized component with $T=2\times 10^6$ K in the nuclear region 
is EM$<6\times 10^{62}$ cm$^{-3}$. This is 30 times smaller than the emission measure 
of the photoionized component (Table ~\ref{tab4}). The upper limit for a hotter collisionally 
ionized component with $T=1\times 10^7$ K is less restrictive and comparable to the 
emission measure of the photoionized Mg {\sc xii} Ly$\alpha$ line: $EM<3\times 10^{63}$ cm$^{-3}$.

The corresponding limits on CI plasma for the NE cloud are EM$<5\times 10^{62}$ and
EM$<1\times 10^{63}$ for $T=2\times 10^6, 1\times 10^7$, respectively. These are
smaller by factors of 15 and 2.5, respectively, than the observed emission measures 
of photoionized O {\sc vii} 2p-1s and Mg {\sc xii} Ly$\alpha$. A CI component would also 
emit thermal bremsstrahlung continuum and strong Fe L lines \citep{skp00}. These are 
more difficult to constrain. 

While we rule out collisional ionization as a major contributor to the spectrum,
a weak component could be present. As we noted above, there is a region in between 
the nuclear region and NE cloud (Fig. ~\ref{fig5}) with a hard spectral color, 
similar to that in the starburst region. This region is relatively free of 
NLR clouds in the {\it HST} [O {\sc iii}] image \citep{mcs94}. With a higher S/N 
{\it Chandra} grating observation at the appropriate roll angle, it should be possible 
to get a spectrum of this region to see if it contains any hot, collisionally 
ionized plasma.  

\subsection{Characterization of the NLR Outflow}

The results of the last two sections show that the X-ray NLR clouds are
ionized and heated primarily by X-ray photons. The luminosity of the 
photoionizing source is consistent with the hidden AGN (Sect. 6.6). There 
is also no indication that the radio jet or lobe is a significant source of X-rays,
otherwise they would appear directly in our X-ray images and spectra. 
Mechanical heating and ionization by shocks are apparently weak, if at all 
present. Any contribution to the nuclear X-ray spectrum from collisionally
ionized gas, either from a starburst or shocked by the radio jet has yet to be
detected.

On the other hand, there is abundant morphological and kinematic evidence 
that the radio jet plays a role in shaping the NLR and that it influences 
cloud dynamics. This leads us to believe that the radio jet interaction with
the NLR is a rather gentle process and indicates a subsonic jet. An analogy 
may be drawn to the radio jets which evacuate bubbles in the IGM surrounding
radio galaxies (e.g., \cite{f00}). 

Our results also indicate that radiation power dominates over kinetic
power output in \object{NGC 1068}. Kinetic power from the radio jet can at 
most provide a small fraction of the heating. However, radiation pressure and 
jet pressure may still be comparable in affecting the dynamics of the NLR. The
kinematics of the NLR suggest two different regions of influence. \cite{ck00}
model the inner $1.5\arcsec$ of the optical [O {\sc iii}] emission line region
with radially accelerating clouds in a hollow bi-cone. This structure may have 
naturally formed by ablation and radiative acceleration of matter from the 
torus. In the region of the NE cloud, the radio lobe may be expanding into the
host galaxy disk \citep{cdg01}. 

We use a grid of XSTAR 2.1 ~\citep{kall01} photoionization models 
covering the parameter space $\log \xi=0.0-4.0$ ($\xi=L/n_\mathrm{e}r^2$) to 
estimate temperatures, ion fractions, and line emissivities in the X-ray NLR. 
We assume solar abundances. The input SED is similar to that used by \cite{kro01}. 
The X-ray spectrum is a power law with photon index $\Gamma=1.8$ and high energy 
cut-off at 100 keV. We assume that each emission line is produced in a region of 
peak emissivity to derive an ionization parameter and temperature for each ion. 

\begin{figure}
\resizebox{\hsize}{!}{\includegraphics{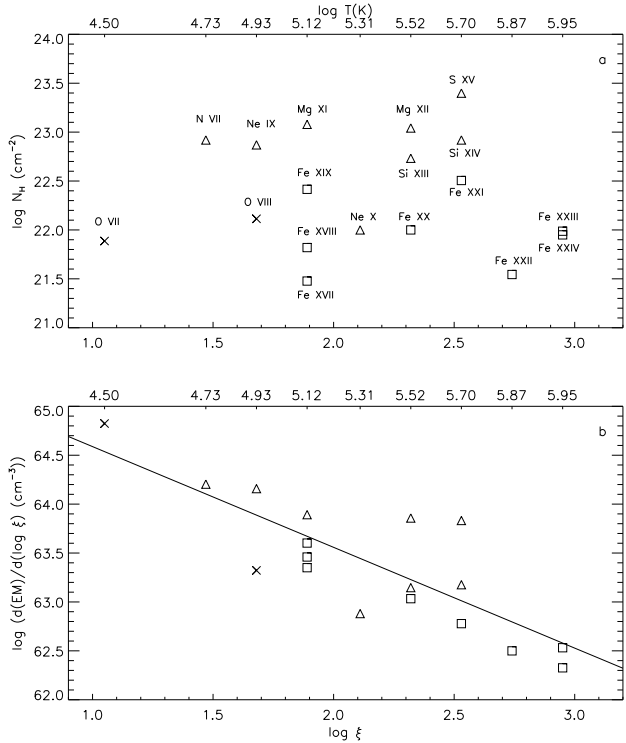}} 
\caption{a) Column density  and b) differential emission measure (DEM) distributions in the 
nuclear region. Ionization parameter $\xi$ and temperature are from XSTAR 
photoionization models, assuming maximum emissivity at solar abundance. The 
distributions cover a large range (2 decades) of ionization, with column densities of similar
magnitude in each ionization state. The relatively low column densities of oxygen (X's) and iron 
(squares) suggest they are under-abundant. The best fit model to the DEM distribution gives $EM 
\sim \xi^{-1}$, indicating equal mass at each ionization parameter.}
\label{fig9}
\end{figure}

Fig. ~\ref{fig9}a shows the distribution of $N_\mathrm{H}$ in the nuclear 
region (from Table ~\ref{tab4}) as a function of $\xi$ and $T$. There is a
large amount of scatter, which we suggest is primarily due to non-solar abundances.
In particular, relatively low oxygen and iron column densities suggest that these 
elements are under-abundant. The distribution covers a large range in ionization 
parameter, corresponding to a factor of 80 in density or a factor of 9 
in distance from the central source. 

We compute the differential emission measure (DEM) distribution (Fig. ~\ref{fig9}b) from
the EM values (Table ~\ref{tab4}) assuming $\mathrm{d}(log \xi)=1.0,0.5,0.25$ for H-like, 
He-like, and Fe L ions, respectively. These $\mathrm{d}(log \xi)$ are estimated from the 
widths of the $i+1$ ion fraction curves $f_{i+1}(\xi)$ computed with XSTAR. Assuming 
equal-size errors for all the data points, we find a best-fit slope of -1.0 and an intercept 
of $log(\mathrm{d}(EM)/\mathrm{d}(\log \xi))=65.6$.
The slope can be understood if we approximate the $N_\mathrm{H}$ distribution by $N_\mathrm{H}(\xi) 
\sim constant$. Then the emission measure drops with ionization parameter, following 
EM$\sim \xi^{-1}$ (Fig.~\ref{fig9}b). Using the definitions of emission measure and ionization 
parameter, this also implies that the product of the mass and X-ray flux is constant, 

\begin{equation}
M(\xi) F_\mathrm{x} \sim n_e V F_\mathrm{x} \sim \xi \times \mathrm{EM} =constant.
\end{equation}

In other words, at any given distance from the source, there are equal masses
of plasma at all observed ionization parameters. This implies a large range of density 
at any given location \citep{ksb02}. \cite{kro01} suggest that a large range in 
density and temperature may result from marginal stability along the vertical branch of 
the cooling curve. It is also apparent from the soft X-ray image (Fig. ~\ref{fig5}) that 
the emission from the X-ray NLR is clumpy and azimuthally asymmetric, with a range of 
colors which isn't a simple function of distance from the nucleus.  

\subsection{The Torus}

The obscuring region in \object{NGC 1068} can be pictured as a collection of
dusty molecular clouds surrounding the nucleus. A conical illumination
pattern for the ionization cone is most simply created by a toroidal ring,
but more complicated geometries may produce a similar effect. The $200\times160$ pc 
(projected) central region of the Chandra image (Fig. ~\ref{fig4}) shows 3-8 keV 
emission from clouds surrounding the hard X-ray peak. This emission may be reflected 
continuum and Fe K$\alpha$ fluorescence. The lack of 1.3-3 keV line emission from this 
region is consistent with a low ionization parameter and high density. 

The extended hard X-ray emission surrounding the nucleus coincides with the 
molecular CO ring measured by \cite{s00}. The two brightest CO spots in this ring 
are located east and southwest of the nucleus and contain an estimated mass of 
$5\times 10^6$ $M_{\sun}$. \cite{s00} model the kinematics of the molecular gas 
with a highly warped disk. The axis of the disk is $70\degr$ to the line of sight 
and has a PA of $20\degr$ at a distance of $1\arcsec$ from the nucleus. 
A warped disk geometry can create a highly irregular X-ray illumination 
pattern, perhaps leading to the observed irregularity in the hard X-ray 
ring. Alternatively, the blobs in the ring may correspond to discrete giant 
molecular clouds. Much closer to the nucleus, H$_\mathrm{2}$O masers outline an 
edge-on portion of the molecular disk with an outer radius of 30 mas (2pc) 
and PA$=4\degr$, and perhaps the inner surface of the torus 
\citep{gbd96,gga96}. It is likely that the column density of the inner disk is
Compton-thick, sufficient to completely obscure the direct line of sight to 
the nucleus. However, the molecular clouds at 100 pc scale also appear to be 
important in defining the geometry of the ionization cone.

A further confirmation of the obscuration model is the reflection spectrum 
from the nuclear region (Fig. ~\ref{fig1}). We first fit the 1.2-5 \AA\ 
(2.5-10 keV) hard X-ray continuum with a PEXRAV \citep{mz95} 
component representing neutral reflection from the molecular torus. We also 
include Gaussian emission lines at the energies of Fe {\sc i} (K$\alpha$ and
K$\beta$) and  Fe {\sc XXV} (r and f), and account for Galactic absorption. 
We use the C-statistic since there are few counts per bin in the HEG 
spectrum. We fit a neutral reflection component with an incident power law 
index of $\Gamma=2.8\pm0.2$. This unusually steep index and significant 
residuals in the 4-5 \AA\ region indicate that an extra continuum component is 
present. Next we add a power law component from ionized reflection. The 
incident power law index is fixed to $\Gamma=1.8$ for both reflection 
components, since it is not well constrained by the data. The best fit 
normalizations at 1 keV for the neutral and ionized reflection components are
$1.2\pm 0.2 \times 10^{-2}$ ph s$^{-1}$ cm$^{-2}$ keV$^{-1}$ and $4\pm 1\times
10^{-4}$ ph s$^{-1}$ cm$^{-2}$ keV$^{-1}$, respectively. The corresponding 
scattered 2-10 keV luminosity from the ionized region is 
$\tau f_\mathrm{c} L_\mathrm{x}=4\pm 1 \times 10^{40} h_\mathrm{75}^{-2}$ 
erg s$^{-1}$.

The equivalent width of the Fe {\sc i} K$\alpha$ line is $1.5\pm 0.3$ keV,
consistent with the value measured by {\it BBXRT} \citep{m93}. Such a
large equivalent width is in agreement with Monte Carlo simulations of neutral
reflection from the inner wall of a Compton-thick molecular torus 
~\citep{ghm94}. It is inconsistent with the smaller equivalent width of 
$\sim0.5$ keV predicted for reflection from the ionized scattering region
\citep{kk87}. 

\subsection{Ionized Scattering Region}

\begin{figure}
\resizebox{\hsize}{!}{\includegraphics{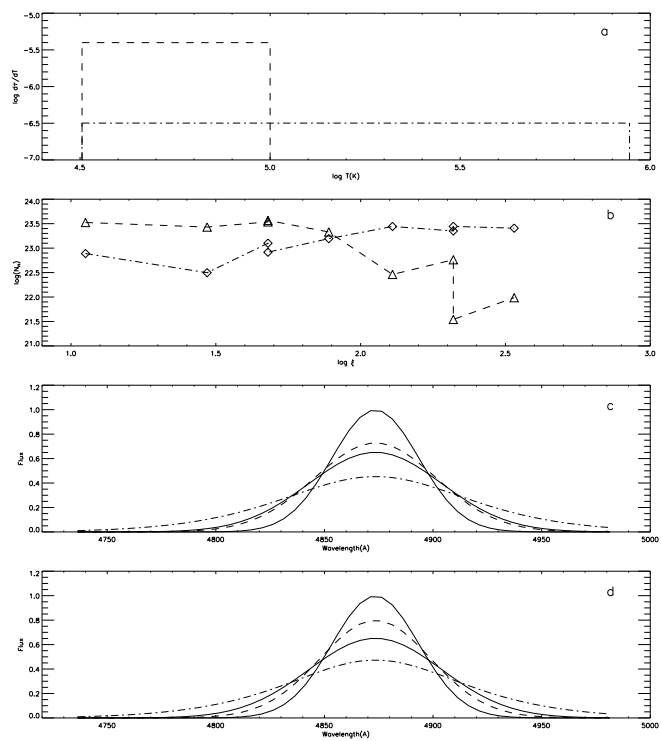}} 
\caption{ a) Model temperature distributions of optical depth in the nuclear
scattering region, assuming two different cut-off temperatures 
($\log T_\mathrm{C}(K)=$5.0,5.95; dash and dot-dash). b) Corresponding
column density distributions, both consistent with the observed distribution
in Fig. 9. c) Scattered H$\beta$ profiles, integrated over the two model temperature 
distributions in panel a. The observed (incident and scattered) line profiles are 
the solid curves \citep{mgm91}. The dashed and and dot-dashed curves show the 
predicted scattered profiles, including  both thermal and turbulent broadening 
(1000 km s$^{-1}$). d) Similar to c), without turbulent broadening.}
\label{fig10}
\end{figure}

An ionized scattering region which gives an indirect view of the hidden 
nucleus is a necessary component of the Seyfert unification theory. 
Ultraviolet light scattered from the inner $4\arcsec$ of \object{NGC 1068}
has a wavelength-independent polarization \citep{ahm94}. This is strong 
evidence for electron scattering as opposed to dust scattering. The optical 
scattering fraction is estimated at $\tau f_\mathrm{c}= 0.015$ from 
spectropolarimetry of the H$\beta$ line \citep{mgm91}. The electron-scattering
region is thought to be associated with the ionized X-ray NLR ~\citep{kk95}. 
As shown in the last section, there is evidence for electron scattering from 
this region in the hard X-ray band. The scattered X-ray continuum and emission
line flux from the nuclear region give estimates of $\tau f_\mathrm{c} 
L_\mathrm{x}$ and $f_\mathrm{c} L_\mathrm{x}$. From the ratio of these, we 
find an electron scattering optical depth of $\tau=0.27 \pm0.08$ (90\% confidence.) 
Combining this result with the optical scattering fraction, the source covering 
fraction in the nuclear region is $f_\mathrm{c}\simeq0.06$, and the luminosity of the 
hidden nucleus is $L_\mathrm{x} \simeq 2.7\times 10^{42}$ erg s$^{-1}$. The 
mean equivalent Thomson scattering depth associated with any given ion ,
$\tau=3.2\times10^{-2}$ (derived from $<N_\mathrm{H}>$, Table ~\ref{tab4}), can account for 
12\% of the scattered X-ray continuum. Integrating over all ionization states,
we can account for the total electron scattering column (Fig. ~\ref{fig10}).

The polarized broad H$\beta$ line from the central $4\arcsec$ of
\object{NGC 1068} appears to be broadened by the thermal motion of scattering 
electrons with mean temperature $<T_\mathrm{e}>=6.7\times 10^4$ K 
\citep{mgm91}. Note that the higher value $<T_\mathrm{e}>=3\times 10^5$ K 
which is often quoted assumes a monochromatic incident line, whereas the lower
value we quote here is for a broad incident line width estimated from the dust
scattered line in the off-nuclear knot. The electron temperature we measure 
for the O {\sc vii} ion ($T_\mathrm{e}=8^{+3}_{-2}\times 10^4$ K) is in good 
agreement with the value from optical spectropolarimetry. The $\sim400$ km s$^{-1}$ 
redshift of the polarized broad H$\beta$ line, attributed to the outflow of the electron 
scattering region ~\citep{mgm91}, is consistent with the mean blue shift of the 
X-ray emission lines, $<v>=-410$ km s$^{-1}$.

A multi-temperature electron scattering region with $\log T(K)=4.5-6.0$ 
is indicated by the observed column density distribution (Fig. ~\ref{fig9}).
We integrate over model temperature distributions (Fig. ~\ref{fig10}a) to predict 
the optical H$\beta$ profile scattered from the X-ray NLR. A simple pedestal function
approximates a flat temperature distribution with a high temperature 
cut-off $T_\mathrm{C}$ and low temperature cutoff at  $\log T(K)=4.5$. The area under the 
curve is normalized to give $\tau=0.27$. We try two different  models, with
$\log T_\mathrm{C}=5.0,5.95$. We integrate the ionic column densities over the
model temperature distributions, and divide by the ionic abundances 
from our XSTAR models to find column density distributions (Fig. ~\ref{fig10}b) 
which are consistent with the observed distribution (Fig. ~\ref{fig9}).

We use a Gaussian incident BLR profile with FWHM$=2900$ km s$^{-1}$ \citep{mgm91}
to compute the scattered  H$\beta$ profile. We also include turbulent broadening of
$b=1000$ km s$^{-1}$, the mean value for the X-ray emission lines (Table 
~\ref{tab3}). The resulting profiles are shown in Fig. ~\ref{fig10}c, and 
compared to the observed polarized broad H$\beta$ line width FWHM$=4400$ km 
s$^{-1}$ \citep{mgm91}. The model with $\log T_\mathrm{C}=5.0$ gives a better
match to the observed H$\beta$ profile than $\log T_\mathrm{C}=5.95$. This 
indicates that there is relatively little plasma at hotter temperatures, or it
would broaden the H$\beta$ line more than is observed. Broadening by turbulent
motions in the NLR has an effect on the profile width for $\log T_C=5.0$, but 
is insignificant for $\log T_\mathrm{C}=5.95$ (Fig. ~\ref{fig10}c,d).

One concern is that the optical-UV bremsstrahlung from the ionized scattering
region not exceed the observed continuum in that wave band. \cite{mgm91}
found that this constraint is satisfied in their models, which assume a
smooth distribution of matter. However, if the scattering clouds are clumpy,
then bremsstrahlung is enhanced by a factor of $<n_\mathrm{e}^2>/
<n_\mathrm{e}>^2$ ~\citep{kk95}. We can make a direct estimate of the optical-UV
bremsstrahlung by integrating over the DEM$(T)$ curve (Fig. ~\ref{fig9}b). 
We predict peak emission of $\nu F_{\nu}= 8.4\times 10^{-14}$ erg s$^{-1}$ cm$^{-2}$
at $\lambda=3000$ \AA. This is 40 times lower than the \cite{mgm91} estimate
of $\nu F_{\nu}<3.4\times 10^{-12}$ erg s$^{-1}$ cm$^{-2}$ (970\AA), which assumed
a radial density distribution of $n_\mathrm{e}\sim r^{-2}$. Hence bremsstrahlung emission
is much weaker than the scattered optical-UV light and will be very difficult
to detect. 

Our Chandra observations confirm the hypothesis that the X-ray NLR and the 
optical/X-ray electron scattering region are one and the same. Both occupy the 
same extended region in the central $3\arcsec$ of the galactic nucleus, and 
have roughly the same optical depth and electron temperature. The expected
variability timescale is the light-crossing time, which is 240 years.
This is apparently at odds with a recent report of 20\% variability in 4 months 
of the scattered X-ray continuum and 60\% variability of the Fe {\sc xxv} (6.7 keV) 
emission line flux ~\citep{cwk02}. However, we can not rule out the possibility 
that the continuum variability comes from a point source which contributes a significant
fraction of the flux. There are not enough photons in our {\it Chandra} observation to 
obtain separate images of the Compton-reflected and electron-scattered components 
of the X-ray continuum. Similarly, we can not spatially distinguish the regions of 
Fe {\sc i} and Fe {\sc xxv} K$\alpha$ emission. It will be important to follow up
with additional high-resolution {\it Chandra} spectroscopy to see if the Fe {\sc xxv} 
line variability is confirmed.

\section{Conclusions}

{\it Chandra} HETGS X-ray spectroscopy and imaging of \object{NGC 1068} 
provide new diagnostics of the spatially resolved narrow-line region and 
support the Seyfert unification model. The X-ray spectrum contains emission 
lines from a large range of ions in the NLR, and fluorescence from the 
molecular torus. The X-ray emitting plasma in the extended NLR is primarily 
photoionized and there is no evidence for a collisionally ionized component. 
Our previous claim of such a component in \object{NGC 4151} \citep{o00} 
was confused by the effects of photoexcitation. 

Observations of the narrow Fe K$\alpha$ line show that it is produced 
predominately in the nuclear region, near the location of the hidden nucleus. 
Its equivalent width is consistent with obscuring torus models. It also appears
that we have detected and resolved the outer regions of the molecular torus in 
hard X-rays. The hard X-ray continuum spectrum from the nuclear region consists 
of a Compton hump from neutral scattering and an electron scattered component. 

We derive ionic column densities, equivalent hydrogen column densities, and
emission measure distributions for both the nuclear region and a cloud $3\arcsec$ 
NE of the nucleus. The column density of the X-ray NLR can account for the entire
electron scattering column in the nuclear region. In addition, its temperature and
velocity are sufficient to cause the observed broadening and redshift of the 
scattered optical broad lines. These facts are all consistent with electron scattering 
in the X-ray NLR providing the view of the hidden Seyfert 1 nucleus.

The results for the X-ray NLR in \object{NGC 1068} may also have important 
implications for the properties of ionized (warm) absorbers seen in the 
spectra of many Seyfert 1 galaxies. Similar velocities and ionization 
parameters are found in warm absorbers and the X-ray NLR. We see from 
{\it Chandra} observations of \object{NGC 1068} that AGN outflows can be 
clumpy and show a wide range of ionization $\log \xi=1.0-3.0$. The differential
emission measure distribution indicates roughly equal mass in each ionization state in
this range. The emission seen in various UV and X-ray lines, though 
occupying roughly the same region, must come from clouds of greatly different 
ionization, temperature, and density. It is important to understand the 
equilibrium among the different phases to develop a complete picture of AGN 
outflows.

\begin{acknowledgements}
This research was partly funded by NASA grant NAG5-7714, NASA HETG grants
NAS8-38249 and NAS8-01129, and Chandra grant GO2-3146X. Observations were 
taken with NASA's {\it Chandra} X-ray Observatory. {\it HST} data were 
obtained from the Multi-mission Archive at the Space Telescope Science 
Institute (MAST). STScI is operated by the Association of Universities for 
Research in Astronomy, Inc., under NASA contract NAS5-26555. Thanks to 
Ski Antonucci, Julian Krolik, Masao Sako, and the referee G. Risaliti for helpful 
discussions and comments. Thanks to Makoto Kishimoto for processing the 
{\it HST} H$\alpha$ image. Thanks to Ming Feng Gu for extensive help with his 
FAC atomic code, and providing Fe L recombination data.
\end{acknowledgements}

\end{document}